\documentstyle[twocolumn,aps,floats,amssymb,epsfig]{revtex}

\begin{document}

\newcommand{\beq}{\begin{equation}}
\newcommand{\eeq}{\end{equation}}
\newcommand{\bea}{\begin{eqnarray*}}
\newcommand{\eea}{\end{eqnarray*}}
\newcommand{\sub}{\scriptsize}
\newcommand{\mat}{\left(\!\!\begin{array}{cc}}
\newcommand{\rix}{\end{array}\!\!\right)}
\newcommand{\ord}{{\cal O}}
\newcommand{\bra}{\langle}
\newcommand{\ket}{\rangle}
\newcommand{\plaq}{\begin{array}{cc}}
\newcommand{\ette}{\end{array}}
\newcommand{\oX}{\overline{X}}
\newcommand{\oY}{\overline{Y}}
\newcommand{\oZ}{\overline{Z}}
\newcommand{\vv}{{\bf v}}
\newcommand{\vt}{{\bf t}}
\newcommand{\vE}{{\bf E}}
\newcommand{\vF}{{\bf F}}
\newcommand{\dh}{\Delta h}
\newcommand{\grad}{\nabla}
\newcommand{\p}{\partial}
\newcommand{\Z}{{\Bbb Z}}
\def\d{{\rm d}}
\def\e{{\rm e}}
\def\i{{\rm i}}
\def\eref#1{(\protect\ref{#1})}
\def\etal{{\it{}et~al.}}

\setcounter{topnumber}{2}
\renewcommand{\topfraction}{0.9}
\renewcommand{\textfraction}{0.1}
\renewcommand{\floatpagefraction}{0.5}

\twocolumn[\hsize\textwidth\columnwidth\hsize\csname @twocolumnfalse\endcsname

\title{Height representation, critical exponents, and ergodicity in\\
the four-state triangular Potts antiferromagnet}
\author{Cristopher Moore and M. E. J. Newman}
\address{Santa Fe Institute, 1399 Hyde Park Road, Santa Fe, NM 87501}
\maketitle

\begin{abstract}
  We study the four-state antiferromagnetic Potts model on the triangular
  lattice.  We show that the model has six types of defects which diffuse
  and annihilate according to certain conservation laws consistent with
  their having a vector-valued topological charge.  Using the properties of
  these defects, we deduce a $2+2$-dimensional height representation for
  the model and hence show that the model is equivalent to the three-state
  Potts antiferromagnet on the Kagom\'e lattice and to bond-coloring models
  on the triangular and hexagonal lattices.  We also calculate critical
  exponents for the ground state ensemble of the model.  We find that the
  exponents governing the spin--spin correlation function and spin
  fluctuations violate the Fisher scaling law because of constraints on
  path length which increase the effective wavelength of the spin operator
  on the height lattice.  We confirm our predictions by extensive Monte
  Carlo simulations of the model using the Wang-Swendsen-Koteck\'y cluster
  algorithm.  Although this algorithm is not ergodic on lattices with
  toroidal boundary conditions, we prove that it is ergodic on lattices
  with free boundary conditions, or more generally on lattices possessing
  no non-contractible loops of infinite order.  To guard against biases
  introduced by lack of ergodicity, we therefore perform our simulations on
  both the torus and the projective plane.
\end{abstract}

\vspace{0.5cm}

]

\section{Introduction}
There has been considerable interest over the last few years in the
properties of classical spin systems possessing highly degenerate ground
states.  Many such models, including ice models\cite{beijeren,knops}, the
triangular Ising antiferromagnet\cite{blote1,nienhuis1}, and dimer
models\cite{zheng,levitov} have been found to have ground state ensembles
which display critical properties such as algebraically decaying spin--spin
correlations and divergent fluctuations in the order parameter.  It is now
known that these properties are associated with the existence of interface
or ``solid-on-solid'' representations for the models, in which sites can be
assigned heights $h$ which vary smoothly over the lattice and which can be
mapped onto the states of the spins or other microscopic variables in a
simple way.  If we assume that this height field behaves as a Gaussian
surface, which is justified if the model is in its rough phase, then the
critical behavior follows in a straightforward
fashion\cite{nijs,nelson,nienhuis3}.  The values of the critical exponents
are related to the stiffness $K$ of the surface and the wavelength of the
appropriate operator on the height lattice.

A few models have been studied for which $h$ is vector rather than scalar
and the ideas above generalize to this case
also\cite{huse,kondev3,raghavan,burton}.  In this paper we look at one
particular model of this type, the four-state antiferromagnetic Potts model
on the triangular lattice.  This turns out to be an especially lucid
example of a model with a vector height, being defined on a simple Bravais
lattice and, as we will show, possessing a very straightforward height
representation.

It is clear that the four-state antiferromagnet does indeed have a highly
degenerate ground state, since the triangular lattice is three-colorable.
By taking a three coloring and introducing a finite density of the fourth
color we can see that the ground state must have an extensive entropy.  In
this paper, we study the properties of the ground state ensemble by
considering first the behavior of defects in the model at zero temperature.
We show that there are six distinct types of non-trivial defects and from
the conservation laws that govern their collisions we deduce that they have
vector charges $\frac{1}{3}\pi$ apart.  We use this observation to derive a
Burgers vector for the model and hence show that when no defects are
present the system has a two-dimensional height representation.  The
defects then correspond to screw dislocations on a $2+2$ dimensional
lattice and we predict that pairs of them will be attracted or repelled
with an entropic Coulomb force proportional to the dot product of their
charges.

We use our height representation to deduce a number of facts about the
four-state model.  First we show that it is equivalent to the three-state
Potts antiferromagnet on the bonds of a hexagonal lattice.  This
equivalence has also been derived using a different approach by
Baxter\cite{baxter1}, but the derivation given here is nonetheless
instructive because it respects the symmetries of the system under
permutation of states in a way that Baxter's does not.  By a simple
geometrical construction we show further that the model is equivalent to
the $q=3$ antiferromagnet on the Kagom\'e lattice, a model which has
previously been studied by Huse and Rutenberg\cite{huse}.  And employing
results due to Kondev and Henley\cite{kondev1}, we show that our model is
equivalent to the fully-packed loop model on the honeycomb lattice with a
loop fugacity of~2.  We generalize these equivalences to several other
cases, including one related to loop models on the square lattice.

The existence of a height representation also implies, as mentioned above,
that the ground state ensemble is critical and we have verified this by
Monte Carlo simulation.  Simulation of this model is not trivial, since no
single-spin-flip algorithm is ergodic and the best-known cluster algorithm,
that of Wang, Swendsen and Koteck\'y\cite{wang1,wang2}, is believed not to
be ergodic under toroidal boundary conditions\cite{SS99} and has not been
proved to be ergodic for any other case.  (For other models, however, it is
known to be ergodic, particularly for models defined on bipartite
lattices\cite{burton,barkema,ferreira}.)  Here we make use of our height
representation to prove for the first time that the algorithm is in fact
ergodic for the $q=4$ Potts model on any lattice with triangular plaquets
satisfying certain topological conditions.  This includes lattices with
free boundary conditions or with the topology of a sphere or a projective
plane.  To take advantage of this result we have performed our simulations
on the projective plane.  The idea of changing the lattice's large-scale
topology in this way to make a sampling algorithm ergodic appears to be
new.

The outline of the paper is as follows.  In Section~\ref{secdefect} we
introduce the model and study the types of defects which occur in it and
their interactions.  In Section~\ref{secheight} we derive the height
representation of the model and thereby demonstrate the model's equivalence
to various others.  In Section~\ref{secenergy} we show how the scaling
exponents characterizing the large-$r$ behavior of correlation functions
can be calculated from the height representation and in
Section~\ref{secforces} we demonstrate the existence of entropic Coulomb
forces between defects.  In Section~\ref{secergo} we describe the Monte
Carlo algorithm we use to simulate the model and in
Section~\ref{secresults} give the results of our simulations.  Finally, in
Section~\ref{secconcs} we give our conclusions.

\section{Defects in the four-state triangular Potts antiferromagnet}
\label{secdefect}
The $q$-state Potts model is a generalization of the Ising model in which a
lattice is populated with spins $s_i$, one on each vertex $i$, which can
take integer values $s_i=1\ldots q$.  The spin states are also sometimes
referred to as ``colors,'' and we will occasionally make use of this
metaphor.  The energy of a configuration is defined to be proportional to
the number of pairs of adjacent sites with the same state
\begin{equation}
H = -J \sum_{\bra ij \ket} \delta_{s_is_j}.
\label{defsh}
\end{equation}
In this paper we consider the Potts model with $q=4$ on the triangular
lattice in two dimensions and with $J<0$ which makes the model
antiferromagnetic so that similarly-colored pairs of adjacent sites are
energetically unfavorable.  We refer to such pairs as ``defects.''  This
model can also be thought of as a discretization of the classical
Heisenberg model in which each site has a three-dimensional unit vector
spin ${\bf S}_i$:
\begin{equation}
H = -J \sum_{\bra ij \ket} {\bf S}_i \cdot {\bf S}_j.
\label{heisenberg}
\end{equation}
This is equivalent to~\eref{defsh} up to a rescaling and an additive
constant if the ${\bf S}_i$ are restricted to the corners of a tetrahedron,
since then the dot product of two spins depends only on whether they are
the same or different.

The ground state entropy per site for this model can be calculated
analytically\cite{baxter2} or closely bounded with series
approximations\cite{schrock}; its exact value is
$3\Gamma\bigl(\frac{1}{3}\bigr)^3/4\pi^2 \simeq 1.460998$.

\begin{figure}
\begin{center}
\psfig{figure=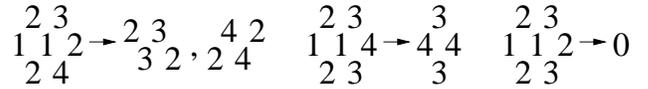,width=3.25in}
\end{center}
\caption{Defects with the same state on both sides of the frustrated bond
we call ``false,'' because they can be annealed away without interacting with
any other defects.}
\label{false}
\end{figure}

Consider the behavior of the model under a single-spin-flip dynamics at
zero temperature.  Such a dynamics allows existing defects to diffuse and
interact in a variety of ways, but creates no new defects.  We now show
that the defects in the model fall into a number of different classes with
well-defined properties.  The simplest
case is that of a defect of the form $\begin{array}{c} b \\
  \begin{array}{cc} a & a \end{array} \\ b
\end{array}$ in which the sites to either side of the defect pair both have
the same state (here denoted $b$).  As Figure~\ref{false} shows, these
defects retain this same form when they move.  Moreover, it is possible for
a defect of this type to disappear entirely if it encounters a neighborhood
of the right configuration.  This is shown in the rightmost portion of the
figure, where the center site in the hexagon can become a~4, resulting in a
defect-free configuration.  Since these defects can be annealed away with
only local moves, and do not require interaction with any other defects, we
call them ``false''; their density can be expected to fall off
exponentially fast in a quench to $T=0$ and so can be ignored where the
long-time relaxation of the system is concerned.

\begin{figure}
\begin{center}
\psfig{figure=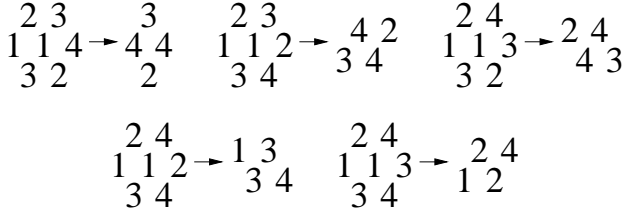,width=3.25in}
\end{center}
\caption{As a defect diffuses, it stays within one of six equivalence
classes, as defined in the text.  Here we show the interaction of a defect
of type $X$ with all possible neighborhoods (excluding symmetry
equivalents).}
\label{diffuse}
\end{figure}

This leaves us with defects $\begin{array}{c} b \\ \begin{array}{cc} a & a
\end{array} \\ c \end{array}$ whose neighboring sites are in different
states $b\ne c$.  As Figure~\ref{diffuse} shows, these defects are
persistent and an isolated defect cannot be annealed away by local moves
alone.  The rest of our analysis in this section will concentrate on these
``true'' defects.

The true defects possess two properties which are conserved both during
diffusion and in interactions with other defects.  The first depends on the
states $a$, $b$ and $c$ which make up the defect and its immediate
neighborhood.  Since $a$, $b$ and $c$ are all different for an isolated
true defect, they divide the four spin states in the model into two pairs,
where $a$ belongs to one pair, and the sites on either side
belong to the other.  In the case of the defect $\begin{array}{c} 2 \\
  \begin{array}{cc} 1 & 1 \end{array} \\ 3 \end{array}$, for example, the
defect sites belong to the pair $\{ 1, 4 \}$ and the adjoining sites to the
pair $\{ 2, 3 \}$.  There are three distinct ways of dividing up the states
in this fashion.

\begin{figure}
\begin{center}
\psfig{figure=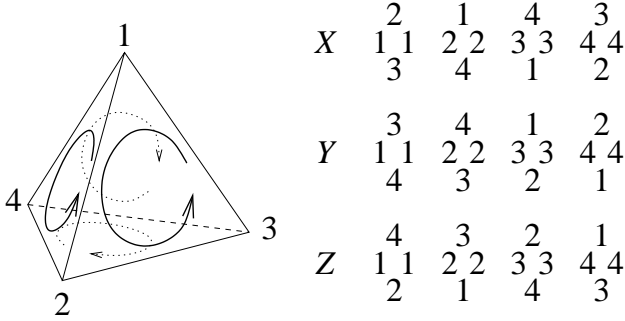,width=3.25in}
\end{center}
\caption{The defect classes $X$, $Y$ and $Z$ all correspond to
counter-clockwise movement on the outside of one of the faces of the
pictured tetrahedron.  The defects $\oX$, $\oY$ and $\oZ$ are their mirror
images and correspond to clockwise movement.}
\label{defects}
\end{figure}

The other conserved property of a true defect is a handedness defined as
follows.  The states $a$, $b$ and $c$, in that order, describe a path which
is either clockwise or counter-clockwise on the outside of one of the faces
of a tetrahedron whose vertices are labelled with the four spin states as
shown in Figure~\ref{defects}.  In order that this property be correctly
conserved we must in addition stipulate that it remains the same under
$120^\circ$ rotations of a defect, but changes sign under $60^\circ$ or
$180^\circ$ ones.  Thus for example the defect $\begin{array}{c} 2 \\
\begin{array}{cc} 1 & 1 \end{array} \\ 3 \end{array}$ considered above has
a counter-clockwise (or positive) handedness, while an inversion or
$60^\circ$ rotation gives us a clockwise (or negative) handedness.

Using these two conserved properties, we can divide the 72 possible defects
into 6 equivalence classes which we label $X$, $Y$, $Z$ and $\oX$, $\oY$,
$\oZ$.  Here the letters denote the pairing of states and the bars (or
absence of them) denote the handedness.  Representative members of the
classes are shown in Figure~\ref{defects}.

\begin{figure}
\begin{center}
\psfig{figure=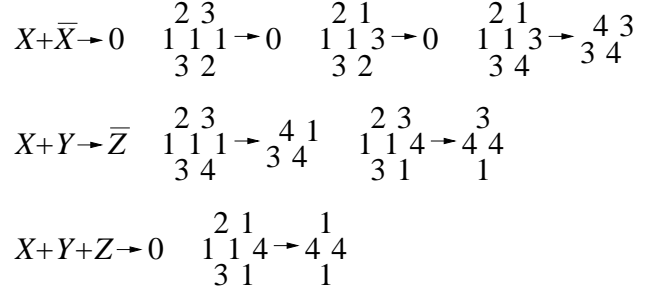,width=3.25in}
\end{center}
\caption{Examples of collisions between defects.  Note that in collisions
such as $X+\oX \to 0$, a false defect is construed as no defect at all.}
\label{coll}
\end{figure}

\begin{figure}
\begin{center}
\psfig{figure=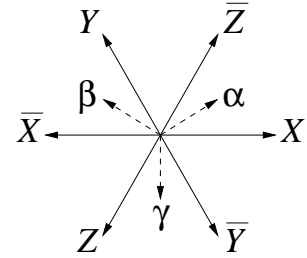,width=1.5in}
\end{center}
\caption{The vector charges assigned to the defects.  The vectors $\alpha$,
$\beta$ and $\gamma$ are defined in the text.}
\label{charge}
\end{figure}

Considering again a single-spin-flip dynamics, we show in Figure~\ref{coll}
a selection of possible collisions between various types of defects.
Although more complex collisions than these can occur, we always find (and
it is proved below) that $X+\oX \to 0$, $X+Y \to \oZ$, and similarly for
cyclic permutations.  If we wish to assign a charge $\chi$ to each particle
such that $\chi(X)+\chi(Y)+\chi(Z)=0$, $\chi(X)=-\chi(\oX)$, and so on, we
can do this with six vectors $\frac13 \pi$ apart as in Figure~\ref{charge}.
We adopt the convention that $|\chi|=1$, $\chi(X)=(1,0)$, and
$\chi(Y)=(-1/2,\sqrt{3}/2)$.  The proof that charge is locally conserved
goes as follows.

It turns out that the total charge within an area can be expressed quite
simply as a sum over the plaquets making up that area.  We start by
writing the charge inside a diamond $\begin{array}{c} a \\
  \begin{array}{cc} b & c \end{array} \\ d \end{array}$ as a sum of
functions of the upward- and downward-pointing triangular plaquets,
$f(a,b,c)+g(d,c,b)$.  Since a $180^\circ$ rotation reverses the charges, we
have $g=-f$, and since a $120^\circ$ rotation keeps them the same, $f$ is
symmetric under cyclic permutations, $f(a,b,c)=f(b,c,a)=f(c,a,b)$.
Furthermore, since a defect-free plaquet has zero charge, $f(a,b,c)=0$ if
$a,b,c$ are all different, and since rotating the tetrahedron of
Figure~\ref{defects} around the corner corresponding to spin state $a$
rotates the charge plane of Figure~\ref{charge} around the origin,
$f(a,a,a)=0$ for any $a$.

This just leaves the case where exactly two of $a$, $b$ and $c$ are equal.
Solving the equations $f(2,1,1)-f(3,1,1)=\chi(X)$,
$f(3,1,1)-f(4,1,1)=\chi(Y)$, and so on, gives
\begin{equation}
f(a,a,b) = \left\{ \begin{array}{ll}
\alpha & \quad\mbox{if $\{a,b\} = \{1,2\}$ or $\{3,4\}$}\\
\beta  & \quad\mbox{if $\{a,b\} = \{1,3\}$ or $\{2,4\}$}\\
\gamma & \quad\mbox{if $\{a,b\} = \{1,4\}$ or $\{2,3\}$}
\end{array} \right.
\end{equation}
where
\begin{eqnarray}
\label{defsalpha}
\alpha &=& \Bigl(\frac{1}{2}, {1\over2\sqrt3}\Bigr),\\
\label{defsbeta}
\beta  &=& \Bigl(-\frac{1}{2}, {1\over2\sqrt3}\Bigr),\\
\label{defsgamma}
\gamma &=& \Bigl(0, -{1\over\sqrt3}\Bigr),
\end{eqnarray}
as shown in Figure~\ref{charge}.

\begin{figure}
\begin{center}
\psfig{figure=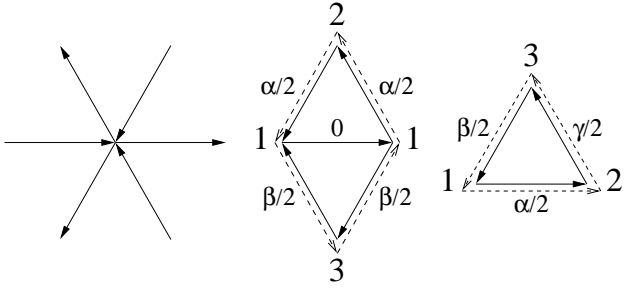,width=3.25in}
\end{center}
\caption{Left: the direction of the vector $\vt$ along the edges of the
  triangular lattice.  Right: the sum $\oint ds \cdot B = \alpha - \beta =
  1$ around a diamond enclosing an $X$ defect and $\alpha/2 + \beta/2 +
  \gamma/2 = 0$ around a defect-free triangle.}
\label{burg}
\end{figure}

If the charge is conserved in all collisions as we have claimed, we should
be able to write it as an integral of some quantity around the perimeter of
the area we are interested in.  It turns out that this is indeed possible
if we define a tensor $B$ on each edge of the lattice equal to the outer
product $B = \vt\otimes\vE$ of two vectors, $\vt$ and $\vE$.  The first of
these points along the edge and gives it a direction as in
Figure~\ref{burg}.  This ensures that $B$ has the necessary change of sign
under $60^\circ$ rotations.  The second is a vector in the charge plane
which depends symmetrically on the states $a$ and $b$ at the two ends of
the edge thus:
\begin{equation}
\vE = \left\{ \begin{array}{ll}
    \alpha/2 & \quad\mbox{if $\{a,b\} = \{1,2\}$ or $\{3,4\}$}\\
    \beta/2  & \quad\mbox{if $\{a,b\} = \{1,3\}$ or $\{2,4\}$}\\
    \gamma/2 & \quad\mbox{if $\{a,b\} = \{1,4\}$ or $\{2,3\}$}\\
    0 & \quad\mbox{if $a = b$}.
\end{array} \right. 
\label{abgeqn}
\end{equation}
Then the charge inside a finite region of the lattice is an integral
around a counter-clockwise perimeter
\begin{equation}
\oint \d{\bf s} \cdot B = \sum ({\Delta{\bf s}} \cdot \vt) \vE.
\label{burgeqn}
\end{equation}
In Figure~\ref{burg}, we show this for a diamond around a single defect and
for a defect-free triangle.  Since all larger regions can be formed by
gluing together diamonds and triangles like these, and since the integral
cancels along the shared edges within the region because of the sign change
imparted on $B$ by $\vt$, it follows that the charge within any region is
correctly given by Equation~\eref{burgeqn}.  This proves our contention
that the charge is conserved in all defect collisions, since the value of
the integral around a line completely enclosing any such collision does not
change when the collision takes place.

\section{Height representation and equivalence to other models}
\label{secheight}
\begin{figure}
\begin{center}
\psfig{figure=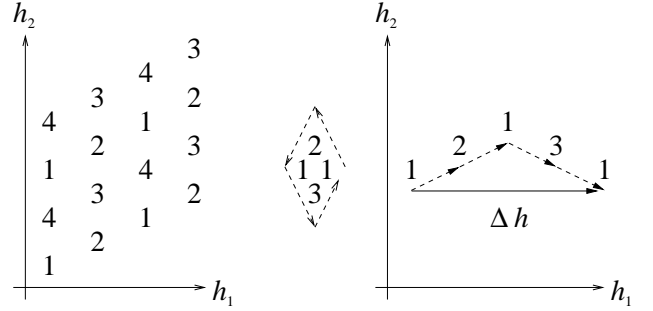,width=3.25in}
\end{center}
\caption{Left: the mapping from heights to states.  Right: integrating
around a defect, in this case an $X$, gives a Burgers vector of $\dh =
\sqrt{3}\,(2\alpha - 2\beta) = 2\sqrt{3}\,\chi$.}
\label{map}
\end{figure}

The integral in Equation~\eref{burgeqn} defines a Burgers vector for a
defect in our model.  Around any defect free region, the Burgers vector is
zero, and hence on a lattice possessing no defects the integral of $B$
between any two points is path independent.  This allows us to define a
height representation for the model in which the height $h$ at any site is
specified uniquely as the integral of $\d{\bf s}\cdot B$ from a single
reference site of known height to the site of interest.  The heights are
thus, like the Burgers vector itself, two-dimensional vectors living on a
triangular lattice with lattice vectors $\alpha$, $\beta$ and $\gamma$,
multiplied by $\pm\sqrt3$ in order to make the lattice constant~1.  In
fact, it is straightforward to show that there is a unique mapping of
heights onto spin states, which is a four-coloring of the height lattice as
shown on the left-hand side of Figure~\ref{map}.  The particular
permutation of the colors in this figure depends on the definition of
$\alpha$, $\beta$ and $\gamma$ given in Equations~\eref{defsalpha}
through~\eref{defsgamma} and on the choice of reference site.

Once we have the height representation for the model, there are a number of
results which follow.  In this section, we use it to demonstrate the
equivalence of the ground state ensemble to a number of other models, some
of which have been studied previously.

\begin{figure}
\begin{center}
\psfig{figure=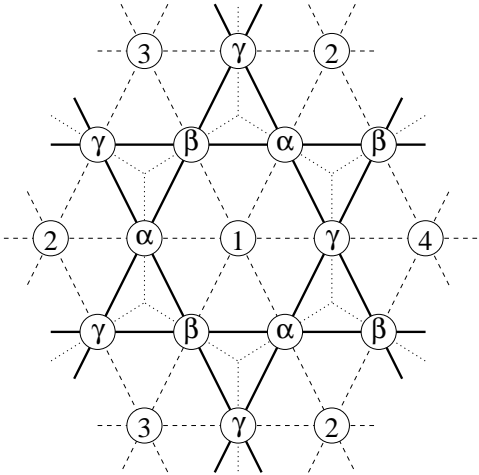,width=2.5in}
\end{center}
\caption{The relationship between the four-state antiferromagnet on the
triangular lattice (dashed) and the three-state antiferromagnet on the
vertices of the Kagom\'e lattice (bold) or the edges of the honeycomb
lattice (dotted).}
\label{kagome}
\end{figure}

First, imagine coloring the edges of the triangular lattice in a
defect-free configuration of the model with three colors $\alpha$, $\beta$
and $\gamma$ according to the height difference along them, as in
Figure~\ref{kagome}.  If we define two edges as neighboring when they bound
the same triangle, then neighboring edges must have different colors since
otherwise two of the vertices of the triangle would have the same spin
state.  Thus the model is equivalent to a three-coloring of the bonds of
the triangular lattice in which no two adjacent bonds are the same color.

The reverse mapping is also possible.  Since $\alpha+\beta+\gamma=0$, the
change in height sums to zero around any triangle, and therefore around any
closed curve, so the height and therefore the state of every site is
well-defined once we have chosen the spin state of one reference site.  As
there are four choices for this reference state, every configuration of
this bond-coloring model corresponds to four ground states of the
four-state triangular Potts model.

Since the edges of a lattice are in one-to-one correspondence with the
edges of its dual lattice, we can also think of the model as a
three-coloring of the edges of the honeycomb lattice, where two edges are
neighbors if they share a vertex (see Figure~\ref{kagome}).  A simple
extension of these mappings is to put a vertex at the midpoint of each edge
on the triangular lattice (or the honeycomb lattice) and connect those
vertices which fall on neighboring edges.  The result, as shown in
Figure~\ref{kagome}, is a Kagom\'e lattice.  (In general, this construction
is called the ``medial graph''\cite{ore}.)  Thus the four-state triangular
Potts antiferromagnet and the three-state one on the Kagom\'e lattice also
have equivalent ground-state ensembles.

A number of these results have appeared previously in one form or another.
Huse and Rutenberg\cite{huse} found a two-dimensional height representation
for the $q=3$ antiferromagnet on the Kagom\'e lattice, which is equivalent
to ours once the equivalence between models demonstrated above is taken
into account.  Baxter\cite{baxter1} (see also Ref.~\onlinecite{bkw})
demonstrated the equivalence of the four-state antiferromagnet and the
bond-coloring model on the hexagonal lattice using an approach somewhat
different from ours.  He defined a cyclic ordering $4\to3\to2\to1\to4$,
drawing arrows from higher states to lower ones (modulo~4) and leaving
edges between states 1 and 3 or 2 and 4 blank.  However, since the
four-state antiferromagnet is invariant under all permutations of the four
states, not just cyclic ones, we feel that the mapping given here better
respects the symmetries of the system under such permutations.

Kondev and Henley\cite{kondev1} showed that the bond-coloring model on the
honeycomb lattice is also equivalent to a fully packed loop (FPL) model on
the honeycomb lattice, where loops are defined as sets of edges alternating
between two of the three colors.  The loops are then contours of the
component of the height perpendicular to the direction corresponding to the
third color\cite{kondev2}.  Each loop can have its colors exchanged without
affecting the surrounding configuration.  Such loops are said to have a
fugacity $n=2$.  (When $n=1$, the FPL model is equivalent to the triangular
Ising antiferromagnet\cite{blote2}.)  An exact solution for the ground
state entropy of the FPL model on the honeycomb lattice for general $n$ has
been given by Batchelor, Suzuki and Yung\cite{batchelor} using a Bethe
ansatz, and differs from Baxter's solution for the entropy of the $q=4$
triangular Potts antiferromagnet\cite{baxter2} by exactly $\log4$, as we
would expect given the equivalence demonstrated above.

The equivalence between four-state antiferromagnets, three-state
bond-coloring models, and fully-packed loop models applies on other
lattices as well.  If a lattice has triangular plaquets and vertices with
even coordination number, we can define an orientation on the bonds with
vectors $\vt$ which go either only clockwise or only counter-clockwise
around each plaquet.  Then we can define $B = \vt \otimes \vE$ as before,
and any path around a defect-free plaquet will have $\dh =
\pm(\alpha+\beta+\gamma) = 0$, so that a consistent definition of $h$
exists.  Furthermore, if we color the bonds with colors $\alpha$, $\beta$
and $\gamma$ then each vertex of the dual lattice has one bond of each
color.  Bonds of any two colors comprise a set of fully packed loops with
fugacity~2, and these are contours of the component of $h$ perpendicular to
the remaining color.

\begin{figure}
\begin{center}
\psfig{figure=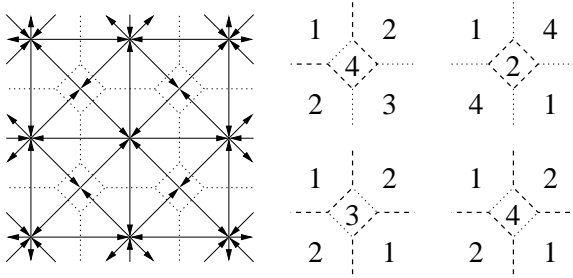,width=3in}
\end{center}
\caption{Left: another lattice on which the four-state antiferromagnet has a
  height representation (lines with arrows), and the corresponding dual
  lattice (dotted).  Right: the fully-packed loops of $\alpha$ and $\beta$
  bonds corresponding to the spin configuration shown.  This model is also
  equivalent to a loop model on the square lattice where loops can collide
  but not cross.}
\label{other}
\end{figure}

As an example, the lattice shown on the left-hand side of
Figure~\ref{other} has vertices of coordination number 4~and 8, and the
orientation of the bonds can be defined as shown.  The $q=4$ Potts
antiferromagnet on this lattice is equivalent to a bond-coloring or
fully-packed loop model on its dual lattice, the truncated square lattice,
whose plaquets are squares and octagons.  If we take the trace over the
sites with four neighbors, we are left with a model on the square lattice
where plaquets of the form ${\sub \plaq a & b \\ c & d \ette}$ are
prohibited, while those of the form ${\sub \plaq a & b \\ b & c \ette}$ and
${\sub \plaq a & b \\ b & a \ette}$ have fugacities 1 and 2 respectively.
This model was studied by Nienhuis\cite{nienhuis2} and is also equivalent
to a model on the square lattice where loops can collide but not cross.  A
similar model where the latter two types of plaquet have equal fugacity was
studied by Burton and Henley\cite{burton}, who found a five-dimensional
height representation for it.

We close this section by defining another model equivalent to the $q=4$
antiferromagnet.  Let us define a chirality on each plaquet of the
triangular lattice according to whether the states at its three corners are
oriented like an interior or exterior face of the tetrahedron in
Figure~\ref{defects}, or equivalently, whether the colors on its edges in
the three-coloring model above cycle clockwise or anticlockwise.  These
plaquets correspond to the vertices of the honeycomb lattice, and it is not
hard to see that 0, 3, or 6 of the vertices of each hexagon must have
positive chirality.  There are 12 configurations of the four-state
triangular antiferromagnet for each state of this model, one for each
choice of the states of two adjacent reference sites.

\section{Free energy and calculation of scaling exponents}
\label{secenergy}
Consider the restricted entropy of the four-state triangular model for a
particular value of some (unspecified) coarse-graining of the height field
$h$.  It is not hard to convince oneself that this entropy is lowest when
$|\grad h|$ is large, and highest for configurations that are
macroscopically flat.  For instance, a three-coloring of the lattice is
flat since any two sites with the same spin state have the same height, and
the set of configurations in the vicinity of such a three-coloring
contributes a large entropy to the ground-state ensemble since every site
has a choice of two colors.  On the other hand, a four-coloring whose
height increases linearly across the lattice corresponds to only one
microstate, since no site has any choices at all.  Building on
considerations such as these we can derive expressions for the scaling
exponents of the model.  Our presentation follows that of Burton and
Henley\cite{burton}.

If the model is in its rough phase, the arguments of the previous paragraph
suggest that it has an effective free energy of the form
\begin{equation}
G = \frac{1}{2} K^{\kappa\lambda} \,\int \grad h_\kappa \,\grad h_\lambda
\,\d x\,\d y,
\end{equation}
where $K^{\kappa\lambda}$ is a stiffness tensor.  However, since the model
is invariant under permutations of the four spin states, and since these
permutations are equivalent to rotations of the height lattice, $K$ must be
a scalar and we have
\begin{equation}
G = \frac{1}{2} K \int \left( |\grad h_1|^2
  + |\grad h_2|^2 \right) \,dx\,dy,
\label{freeeqn}
\end{equation}
where $h_1$ and $h_2$ are components of the height field along any two
perpendicular directions in the height space.  (In our calculations we have
taken $h_1$ and $h_2$ along the directions indicated in Figure~\ref{map}.)
In frequency space this free energy decouples into a sum over independent
Gaussians, giving spatial correlations between the heights at points a
distance $r=|{\bf r}_2 - {\bf r}_1|$ apart of
\begin{equation}
\bra |h({\bf r}_2)-h({\bf r}_1)|^2 \ket \simeq \frac{1}{\pi K} \log r + C
\label{hcorr}
\end{equation}
for large $r$, where $C$ is a constant.

Quantities such as spin and local magnetization are periodic functions
$f(h)$ of the height and hence can be Fourier expanded in the height.  We
can calculate the spatial correlations in any one such Fourier component
$f_g\propto \e^{\i g \cdot h}$, having frequency $g$ in height space, using
the fact that $h$'s Fourier components are Gaussianly distributed.  This
gives us
\begin{eqnarray}
\bra f_g({\bf r}_1) f_g({\bf r}_2) \ket &\propto&
\bigl\bra \e^{\i g \cdot [h({\bf r}_2)-h({\bf r}_1)]} \bigr\ket\nonumber\\
&=& \e^{-\frac{1}{2} g^2 \bra |h({\bf r}_2)-h({\bf r}_1)|^2 \ket}
  \sim r^{-(d-2+\eta)},
\end{eqnarray}
for large $r$ where $\eta$ is the anomalous dimension of the correlation
function and $d$ is the dimensionality of the lattice.  Given that $d=2$ in
the present case and making use of Equation~\eref{hcorr} we then find that
\begin{equation}
\eta = \frac{g^2}{2\pi K}.
\end{equation}
If a quantity has several non-zero Fourier components, then the one with the 
smallest $\eta$---i.e.,~longest wavelength---will dominate for large $r$.  

If $K$ is not a scalar, these equations generalize to
\begin{equation}
\bra (h'_\kappa - h_\kappa)(h'_\lambda - h_\lambda) \ket
  = \frac{K^{-1}_{\kappa\lambda}}{\pi} \log r + C_{\kappa\lambda}
\end{equation}
and
\begin{equation}
\eta = \frac{g K^{-1} g^\dagger}{2\pi}.
\label{etaeqn}
\end{equation}
For instance, transforming a scalar $K$ to triangular coordinates where
basis vectors are $\frac{2}{3}\pi$ apart rather than orthogonal gives a
matrix $K'$
\begin{equation} K' = K \mat 1 & -1/2 \\ -1/2 & 1 \rix \mbox{ and } 
   (K')^{-1} = \frac{1}{K} \mat 4/3 & 2/3 \\ 2/3 & 4/3 \rix.
\label{kprime}
\end{equation}

Using the equivalence of the bond-coloring model to the fully-packed loop
model on the honeycomb lattice, and relating the vortex--antivortex
correlation function to the probability that two sites lie on the same
loop, Kondev and Henley\cite{kondev4} have shown that these models are
exactly at their roughening transition and have a stiffness of
$K=\frac{2}{3} \pi$.  Since our four-state model is equivalent to these
models, it has the same value of $K$.  We will use these results below to
calculate the scaling exponents of various quantities for comparison with
Monte Carlo experiments.

\section{Forces between defects}
\label{secforces}
Since the energy of a pair of defects is 2 regardless of how far apart they
are, there is no energy gradient to drive a force between them.  However,
there is an entropic force, driven by the fact that the presence of a free
defect reduces the entropy within an area of radius $r$ by an amount
proportional to $\log r$.

In a model with a one-dimensional height representation, a defect with
Burgers vector $b$ has an average field around it
\begin{equation}
|\grad h| = \frac{b}{2\pi r},
\end{equation}
giving a force between two defects with Burgers vectors $b$ and $b'$
\begin{equation}
F = \frac{K}{\pi} \frac{bb'}{r}.
\end{equation}
When coupled with a mobility $\Gamma$, this gives an average velocity
to the defects of
\begin{equation}
v = \Gamma F = \frac{\Gamma K}{\pi} \frac{bb'}{r}.
\end{equation}
Such forces have been measured numerically for the three-state Potts model
on the square lattice\cite{moore}.  The generalization to a
higher-dimensional height representation is straightforward.  Since the
free energy in Equation~\eref{freeeqn} is a sum of independent terms in
$h_1$ and $h_2$, the force between two defects with Burgers vectors ${\bf
  b}$ and ${\bf b}'$ will be proportional to $b_1 b'_1 + b_2 b'_2 = {\bf b}
\cdot {\bf b}'$.  Since the Burgers vector for a defect is equal to
$2\sqrt{3}$ times the charge $\chi$ on that defect (see
Section~\ref{secheight}), this gives us a force of
\begin{equation}
F = \frac{12 K}{\pi} \frac{\chi \cdot \chi'}{r}.
\end{equation}
In other words, an $X$ and a $Y$ will be attracted to each other, but only
half as strongly as an $X$ and an $\oX$, and an $X$ and a $\oY$ will be
repelled half as strongly as an $X$ and another $X$.

\section{Monte Carlo simulation}
\label{secergo}
In order to simulate correctly the properties of the ground-state ensemble
of a system, it suffices to find a set of update moves which take us from
one ground state to another without introducing any defects, such that
every ground state can be reached from every other in a finite number of
moves on a finite lattice.  An algorithm based on such a set of moves is
said to be ergodic, and it can be shown that any ergodic algorithm will
sample all ground states with equal frequency over the course of a long
simulation (see Ref.~\onlinecite{NB99} for example).  Unfortunately, it
turns out to be quite difficult to find a suitable ergodic set of moves for
the four-state Potts model considered here.  To begin with, it is clear
that no single-spin-flip dynamics can be ergodic because finite defect-free
regions can be pinned under such a dynamics, and remain pinned no matter
what happens outside them.  For instance, in the hexagon ${\sub
\begin{array}{c}
\begin{array}{cc} 1 & 2 \end{array} \\
\begin{array}{ccc} 3 & 4 & 3 \end{array} \\
\begin{array}{cc} 2 & 1 \end{array}
\end{array}}$
every spin has at least one neighbor of each of the other states, so no
spin can change state.  Since there is a positive density of such clusters
in a random ground state on a large lattice, single-spin-flip dynamics will
only explore an exponentially small fraction of the possible
configurations.

So we are forced to turn to a cluster update algorithm to simulate this
model.  The algorithm we use in this paper is the zero-temperature limit of
the Wang-Swendsen-Koteck\'y (WSK) cluster algorithm\cite{wang1,wang2} for
Potts antiferromagnets, which is defined as follows.  At each step in the
simulation we choose two of the four colors on the lattice, identify all
connected clusters containing only these two, and, in each cluster
independently, either switch the two colors or leave them untouched with
probability $\frac{1}{2}$.  Clearly this preserves the property that
neighbors have differing states, and, as we will show, it is
computationally quite efficient.

At finite temperature the WSK algorithm is trivially ergodic, as
demonstrated by Wang~\etal\ in their original paper.  With a little more
effort it is also possible to prove that it is ergodic at $T=0$ for all
$q>2$ on the square lattice\cite{barkema} or more generally on bipartite
lattices\cite{burton,ferreira}.  The triangular lattice however is not
bipartite, and in fact the algorithm is known not to be ergodic at $T=0$ on
triangular lattices with toroidal boundary conditions, at least for certain
lattice dimensions\cite{SS99}.  In this section we make use of our height
representation to prove that the algorithm is ergodic for certain other
types of boundary conditions.

Previous proofs of the ergodicity of the WSK algorithm at zero temperature
have relied on defining a specific ``target configuration'' of the spins on
the lattice and demonstrating that any given configuration can be
transformed into this one in a finite number of reversible moves.  For
bipartite lattices the target configurations used have been checkerboard
colorings.  Here we use the same approach to prove the algorithm ergodic on
the triangular lattice but, since the lattice is not bipartite and
therefore does not permit checkerboard colorings, we use instead a
three-coloring of the lattice as our target configuration.

\begin{figure}
\begin{center}
\psfig{figure=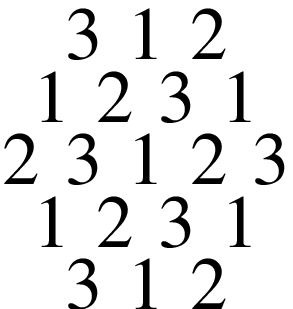,width=0.75in}
\end{center}
\caption{A three-coloring of the triangular lattice.  There are six such
  colorings for each choice of three colors.}
\label{threecol}
\end{figure}

There are six possible three-colorings of the triangular lattice for each
choice of 3 of the 4 colors.  We illustrate one of them in
Figure~\ref{threecol}.  We define a {\em domain\/} to be a connected set of
sites whose colors coincide with one of these three-colorings.  The three
colors in such a domain fall on three triangular sublattices with lattice
parameter $\sqrt3$ times that of the fundamental lattice.  Our goal is to
add new sites to the domain, singly or in groups, until the domain fills
the entire lattice.  If we can do this with reversible cluster moves from
any given starting state, then our algorithm must be ergodic, since we can
get from any state to the target and then from the target to any other
state.  (It makes no difference which three-coloring we take as our target
configuration since we can get from any one to any other in at most three
steps of the WSK algorithm---one to get the three colors right and at
most two to put the colors on the correct sublattices.)

\begin{figure}
\begin{center}
\psfig{figure=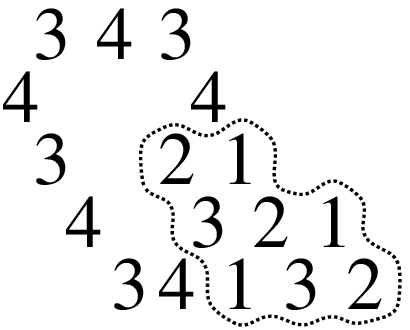,width=1.25in}
\end{center}
\caption{A configuration which defeats the WSK algorithm.  We would like to
  grow the domain indicated by the dotted line by adding to it the spin
  with state~4 immediately above it and propose to do this by changing each
  of the 3s in the domain into a 4.  But a path of 3s and 4s connects the
  domain to the new site, so it will be changed to a~3 when we do this.  In
  the text, we show that configurations of this kind are forbidden within
  the ground-state ensemble because any contractible loop of this kind must
  contain a defect.}
\label{path}
\end{figure}

Our approach is as follows.  We choose a site which lies outside our
domain, but is adjacent to it.  The color of this site, call it $a$,
differs from the color $b$ of the sites inside the domain on the same
sublattice.  If we can switch the colors $a$ and $b$ everywhere inside the
domain, while leaving the color of the new site unchanged, it will now
match the other sites on that sublattice and we will have added it to the
domain.  (Note that the domain now has a new three-coloring, resulting from
our switching $a$ and $b$.)  This can be done trivially in a single Monte
Carlo move if we make the right choice of two colors for the move, but
there is a catch.  The problem is that there may be some cluster connecting
the new site to a (possibly quite distant) site in the domain via sites of
colors $a$ and $b$.  If such a cluster exists, then the new site will get
flipped whenever we change colors within the domain, so that its color will
always differ from that of the sites on the same sublattice in the domain.
An example of such a situation is shown in Figure~\ref{path}.

Happily, we can invoke the height representation to prove that any state in
which such a path exists must contain at least one defect, and hence that
such paths cannot exist in a ground state of the system.  To see this we
consider a closed loop of sites formed by the path outside the domain,
completed with any path of our choice within the domain.  To make things
concrete, let us suppose that $a$ and $b$ are the states 4 and 3 as in
Figure~\ref{path}.  The portion of our closed path outside the domain
consists by definition of sites of only these two colors and hence the
change in height $\dh$ from one end to the other must be a multiple of, in
this case, $\alpha$.  Within the domain, on the other hand, the heights on
all sites belonging to a particular sublattice are the same, so that $\dh$
for the portion inside the domain is just equal to the change in height
resulting when we change a 3 in the domain to a 4, or $\beta - \gamma$ in
this case, which is perpendicular to $\alpha$.  In general, $\dh$ for
changing some site in a three-coloring from one color to another is
perpendicular to the $\dh$ between neighboring sites of those colors.  This
means that the sum of the $\dh$s for the two parts of the loop cannot be
zero, and hence the Burgers vector around the loop is non-zero and the loop
must contain a defect.

\begin{figure}
\begin{center}
\psfig{figure=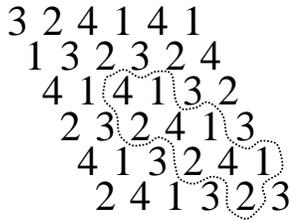,width=1.5in}
\end{center}
\caption{A configuration on the torus with nonzero Burgers vector around
  its non-contractible loops.  The domain denoted by the line cannot grow
  monotonically under the WSK algorithm.}
\label{domain}
\end{figure}

Unfortunately, this does not quite prove the ergodicity of the algorithm.
Certainly a configuration like that in Figure~\ref{path} can be ruled out,
because there must be at least one defect within the loop, and hence the
lattice cannot be in a ground state.  (The reader might like to populate
the interior of the loop with spins just to check that this is indeed
true.)  However, there is another possibility.  If we have some form of
periodic boundary condition on our lattice, it may be possible for the loop
to go off one side of the lattice and come back on another and rejoin the
domain that way.  It turns out that such a loop can have a non-zero Burgers
vector even when the lattice is in a ground state.  In essence, the lattice
possesses a non-localized defect, without there being a defect anywhere in
particular.  The crucial difference between the two types of loops is that
in the first case the loop is {\em contractible,} meaning that it can be
shrunk to a point by shifting it one plaquet at a time.  For a loop of this
type, the situation depicted in Figure~\ref{path} applies, and the
arguments given above are correct.  If however the loop is
non-contractible, then it is possible for there to be a non-localized
defect and we cannot prove the ergodicity of the algorithm.  To give an
example, we show in Figure~\ref{domain} a configuration of the model on a
lattice with toroidal boundary conditions.  This lattice has two
fundamental non-contractible loops, one wrapping around the boundary
conditions horizontally, and one vertically.  For the configuration shown,
these two loops have Burgers vectors of $2\beta + 4\gamma$ and $4\beta +
2\gamma$ respectively, even though there are no localized defects.  In
fact, the domain inside the dotted line shows how the algorithm can fail.
Any attempt to add a new site to the domain by exchanging two colors inside
the domain will change the color of the new site as well, because the new
site is connected to the domain by paths which wrap around the boundary
conditions.  This does not actually prove that the WSK algorithm is {\em
  not\/} ergodic on this lattice, only that we cannot prove it to be so
using arguments of the type given here in which domains gain sites but do
not lose them.  However, Salas and Sokal\cite{SS99} report that they found
a configuration on a $6\times6$ torus which can only be transformed into a
small number of others, thus showing that the algorithm is not in general
ergodic on the torus.  Huse and Rutenberg, in their study of the
three-state model on the Kagom\'e lattice, noted a similar lack of
ergodicity in a loop-flipping algorithm on the torus\cite{huse}.  It seems,
therefore, that it would be imprudent to conduct simulations solely on a
toroidal or similar lattice.

\begin{figure}
\begin{center}
\psfig{figure=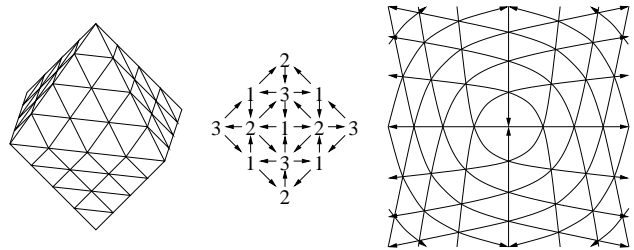,width=3.25in}
\end{center}
\caption{Left: an octahedral lattice with $L=4$.  The WSK algorithm is
  ergodic on this lattice.  Center: a three-coloring of this lattice around
  one of the vertices of the octahedron and the definition of the
  orientation vector $\vt$ along the bonds.  Right: the structure is that
  of a triangular lattice everywhere except at the vertices of the
  octahedron.}
\label{octa}
\end{figure}

On lattices with no non-contractible loops, however, the algorithm is, by
the arguments given above, ergodic.  Examples of such lattices include the
infinite lattice and any finite lattice with free boundary conditions,
where opposite edges of the lattice have no connection to each other.
Unfortunately, the first of these is impractical for computer simulations,
and the second suffers rather dramatically from finite-size effects.  A
better solution is to perform our simulation on a lattice with periodic
boundary conditions, but with a topology chosen so that there are no
non-contractible loops.  The simplest example is the sphere.  Because its
Euler characteristic is~2, there is no way to cover a sphere with a regular
triangular lattice; some sites have to have less than six nearest
neighbors.  Fortunately, our proof works not just for the triangular
lattice, but for any lattice where both the height representation and the
target three-coloring can be defined.  As we showed in the previous
section, the height representation can be defined for any planar lattice
with triangular plaquets and even coordination number, so that we can place
vectors $\vt$ on the bonds oriented so that they go only clockwise or only
counter-clockwise around each plaquet.  We can also define a three-coloring
on such a lattice by having colors cycle upward or downward along $\vt$.
Thus, we can for instance perform a simulation on a lattice tiling the
surface of an octahedron where 6 sites have four nearest neighbors, and the
WSK algorithm will be ergodic.  In Figure~\ref{octa} we depict one such
lattice and illustrate how the three-coloring behaves near its corners.

\begin{figure}
\begin{center}
\psfig{figure=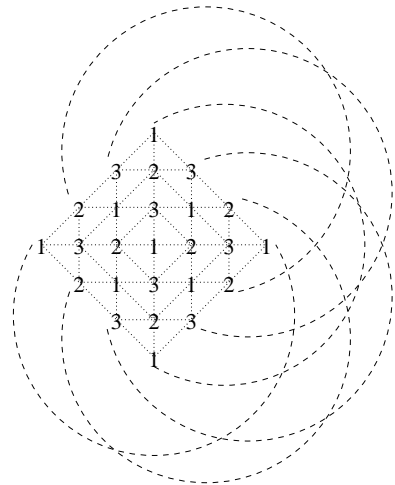,width=2in}
\end{center}
\caption{An $L=3$ lattice whose topology is that of a projective plane.
  Pairs of sites connected by dotted lines are identified.  Note the
  consistency of the three-coloring.}
\label{proj}
\end{figure}

Another possible topology is the projective plane---a hemisphere with
diametrically opposing points on the equator identified.  To create a
lattice with this topology, we simply take the upper half of the octahedron
depicted in Figure~\ref{octa} and identify sites along its square equator
as shown in Figure~\ref{proj}.  Since the projective plane has only half as
much curvature as the sphere (its Euler characteristic is~1) there are now
just three sites with four neighbors rather than six.  Thus this lattice is
proportionately flatter than the sphere with the same number of sites.
Note that the three-coloring is well-defined for any lattice size.

Although the projective plane does possess non-contractible loops, every
such loop has the property that going around it twice makes it
contractible.  Formally, the fundamental group $\Pi_1$ for the space is
$\Z_2$, the integers mod~2.  This means that the Burgers vector of any such
loop must satisfy $\dh + \dh = 0$, so $\dh = 0$.  (Kawamura and
Miyashita\cite{kawamura} show that the Heisenberg antiferromagnet on the
triangular lattice has defects with a $\Z_2$ charge, so there are spin
systems in which $\dh$ can be nonzero even though $2 \dh = 0$.  This
however is not the case with the present system.)  Thus non-contractible
loops are tolerable as long as they have finite order, i.e.,~as long as
their $n^{\rm th}$ multiple is contractible for some finite $n$.  The
sphere and projective plane are the only finite two-dimensional manifolds
satisfying this condition, since both tori and Klein bottles have
non-contractible loops of infinite order.

While Monte Carlo calculations performed on spheres and projective planes
may seem outlandish, they are important in the present case in order to
rule out the possibility that lack of ergodicity is introducing a bias into
our results.  The drawback is that we are forced to introduce a small
number of atypical sites into the lattice (those with only four neighbors)
which, for example, makes calculation of spatial correlation functions more
difficult.  In this paper we strike a compromise by performing some
(probably non-ergodic) simulations on toroidal lattices, which give
excellent statistics for correlation functions and other quantities, and
some on the projective plane, which give poorer statistics, but are more
trustworthy.  In fact, we find that there are no physical measurements for
which the two topologies disagree, so it is possible that the simulations
on the torus are ``sufficiently ergodic'' for our purposes, although we
cannot guarantee this.

Just as with the height representation, our proof of ergodicity will work
for other lattices as well, whenever we can give each bond a direction such
that the bonds around each triangular plaquet go either only clockwise or
only counter-clockwise.  It can also be used for some lattices which have a
mixture of ferromagnetic and antiferromagnetic bonds, but which have no
frustration.  At $T=0$ vertices with a ferromagnetic bond between them can
simply be identified, since they must have the same state.  Then the WSK
algorithm is ergodic on such lattices if the resulting graph fits the
conditions above.

\section{Results of Monte Carlo simulations}
\label{secresults}
\begin{figure}
\begin{center}
\psfig{figure=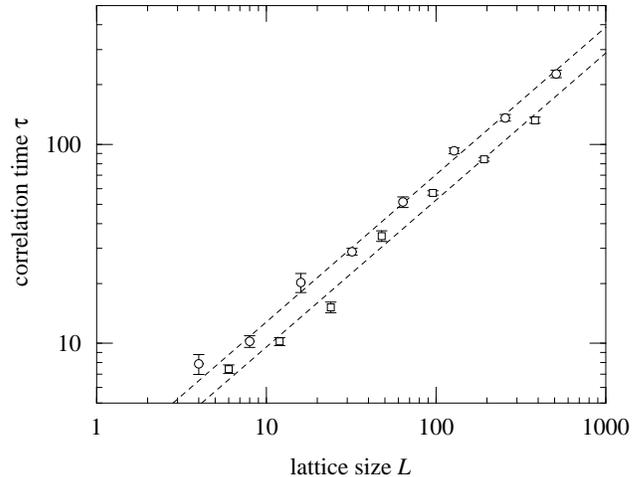,width=3.25in}
\end{center}
\caption{Correlation times for the WSK algorithm on the projective plane
  (circles) and the torus (squares) measured in Monte Carlo steps.  Both
  lattices give a dynamic exponent of $z = 0.74 \pm 0.02$.}
\label{tau}
\end{figure}

We have performed extensive simulations of the four-state triangular model
using the WSK algorithm on lattices having the topology both of the
projective plane (for which the algorithm is definitely ergodic) and of the
torus (for which it probably is not).  On the projective plane we simulated
systems with linear dimension $L$ equal to a power of two from $L=4$ up to
$L=1024$.  On the torus the three-coloring of the lattice is only well
defined for $L$ a multiple of three, so we simulated systems with $L=6$,
$12$, $24\ldots768$.  In each case simulations ran up to one million Monte
Carlo steps for the largest lattices.  In order to allow accurate error
estimation, and also to examine the efficiency of the algorithm, we first
measured the (exponential) correlation time $\tau$ in Monte Carlo steps as
a function of system size.  Figure~\ref{tau} shows the results for both
topologies.  In each case the dynamic exponent $z$, defined by $\tau\sim
L^z$, takes the value $0.74\pm0.02$, which is comparable with values for
cluster algorithms for ferromagnetic Potts models\cite{CB91}.  Correlation
times for the projective plane are about a factor of $1.3$ larger than
those for the torus with the same value of~$L$, which is presumably because
the projective plane has more sites on it.  (There are $L^2$ sites on a
torus of linear dimension~$L$, but $2L^2+1$ on the projective plane.)

In practical terms, since a single step of the WSK algorithm updates a
number of spins which scales with the area $L^d$ of the lattice, the CPU
time taken to generate a given number of independent lattice configurations
scales as $L^{d+z} \simeq L^{11/4}$.

In order to measure the critical exponents for the model, we chose two
different definitions for the order parameter and measured two-point
correlations and fluctuations for each of them.  From these results we
extracted values for the correlation exponent $\eta$ by direct measurement,
and for the susceptibility exponent $\gamma/\nu$ by finite size scaling.
These exponents are not independent; we expect them to be related by the
Fisher scaling law $\gamma/\nu = 2 - \eta$\cite{fisher,binney}.  Our two
order parameters were defined as follows:
\begin{enumerate}
\item The simplest magnetization measure is just $m_k = N_k - \frac14 N$
  where $N_k$ is the number of spins on the lattice in spin state $k$.  The
  two-point connected correlation for this magnetization, averaged over
  $k$, is then $G_c(i,j) = \delta_{s_is_j} - \frac14$ and the
  susceptibility is $\chi = \sum_k m_k^2$.
\item We also examined the ``staggered magnetization'' defined by Huse and
  Rutenberg\cite{huse} for the three-coloring model on the hexagonal
  lattice, to allow direct comparison with their results.  In the context
  of the present model, this magnetization may be thought of as a complex
  number $\sum_l c_l r_l$ where the sum is over bonds on the triangular
  lattice, $c_l$ represents the color $\alpha$, $\beta$ or $\gamma$ of bond
  $l$ as defined in Equation~\eref{abgeqn}, and $r_l$ is a static reference
  coloring corresponding to the color of bond $l$ in any of the 24 possible
  three-colorings of the lattice.  The staggered magnetization $m_i$ on a
  site $i$ can be defined as the sum over the bonds $l$ connected to that
  site and the two-point correlation is then given by $G_c(i,j) = m_i
  m_j^*$ and the susceptibility by $\chi = \sum_i m_i m_i^*$.
\end{enumerate}

\begin{figure}
\begin{center}
\psfig{figure=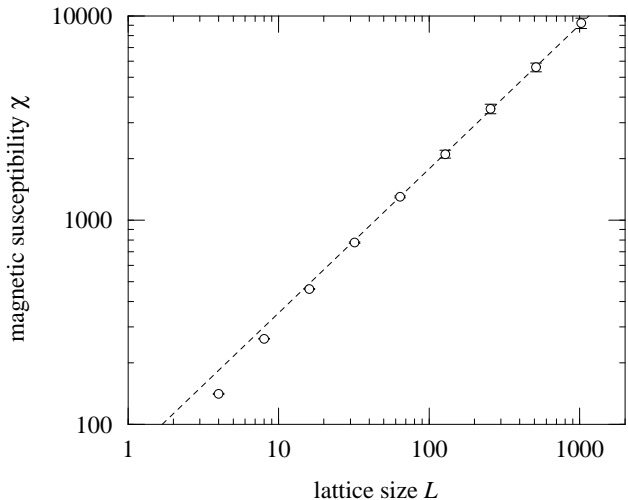,width=3.25in}
\end{center}
\caption{The susceptibility for the staggered magnetization on the projective
  plane as a function of lattice size $L$.  The dotted line is a fit to the
  last five points and gives $\gamma/\nu = 0.71 \pm 0.02$.  Theory predicts
  $\gamma/\nu = \frac23$.}
\label{stagsusc}
\end{figure}

\begin{figure}[t]
\begin{center}
\psfig{figure=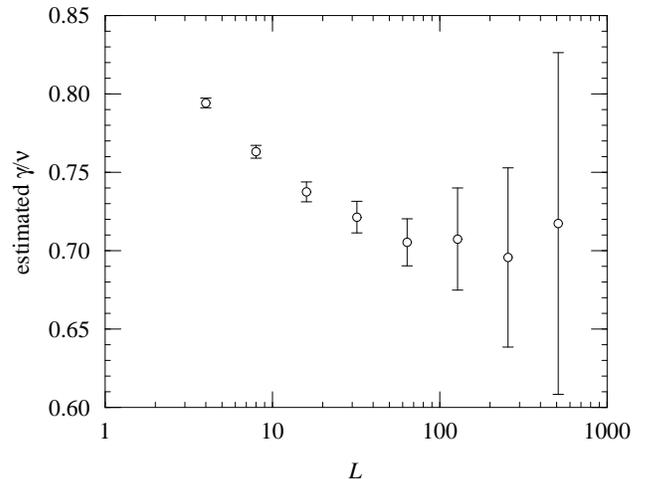,width=3.25in}
\end{center}
\caption{Estimates of $\gamma/\nu$ obtained from fits to simulation
  results for the staggered magnetic susceptibility on lattices of size
  ranging from $L$ up to $1024$, as a function of $L$.  These tend downward
  with increasing $L$, presumably as a result of logarithmic corrections to
  the scaling form.}
\label{downward}
\end{figure}

The staggered magnetization is in fact slightly the easier of these two to
analyze, so we examine this case first.  Huse and Rutenberg\cite{huse}
pointed out that the staggered magnetization has a wavelength $\sqrt{3}$ on
the height lattice.  Using Equation~\eref{kprime} to transform from
Cartesian coordinates to triangular ones, Equation~\eref{etaeqn} then gives
$\eta = \frac{4}{3}$ and $\gamma/\nu = \frac{2}{3}$.  Figure~\ref{stagsusc}
shows our simulation results for the susceptibility on the projective
plane.  A least-squares fit gives $\gamma/\nu = 0.71\pm0.02$, which is a
little greater than the expected value, but, as Figure~\ref{downward}
shows, there is a clear downward trend in the value with increasing system
size.  Huse and Rutenberg saw similar behavior in their bond-coloring model
(in fact their value for $\gamma/\nu$ is almost exactly the same as ours)
and they attributed it to logarithmic corrections to the scaling forms
which arise because the height representation is at its roughening
transition (see Section~\ref{secenergy}).

\begin{figure}
\begin{center}
\psfig{figure=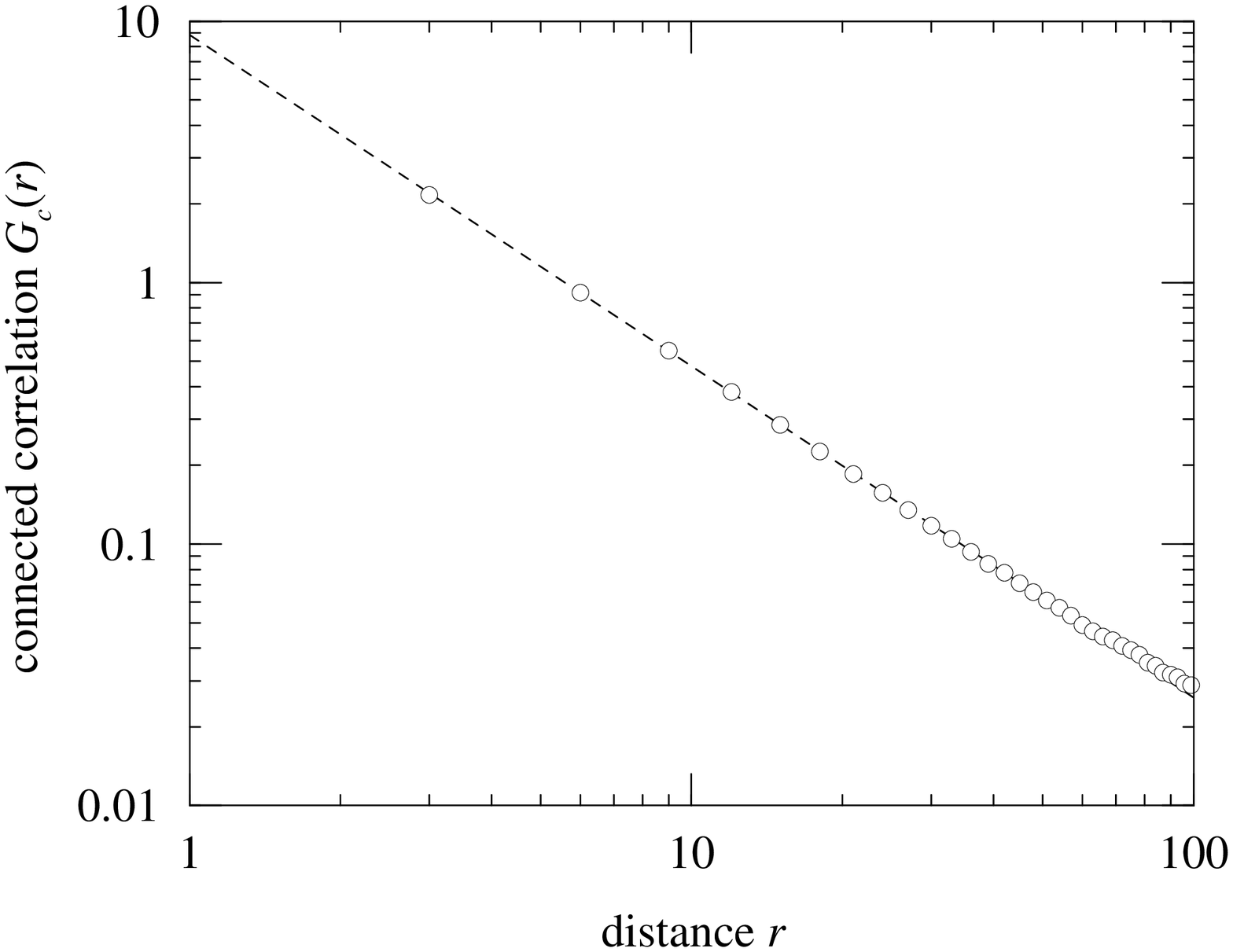,width=3.25in}
\end{center}
\caption{The connected correlation $G_c(r)$ for the staggered magnetization
  on the torus.  The best power-law fit is indicated by the dotted line and
  gives $\eta = 1.27 \pm 0.01$.  Theory predicts $\eta = \frac43$.}
\label{stagtorus}
\end{figure}

\begin{figure}
\begin{center}
\psfig{figure=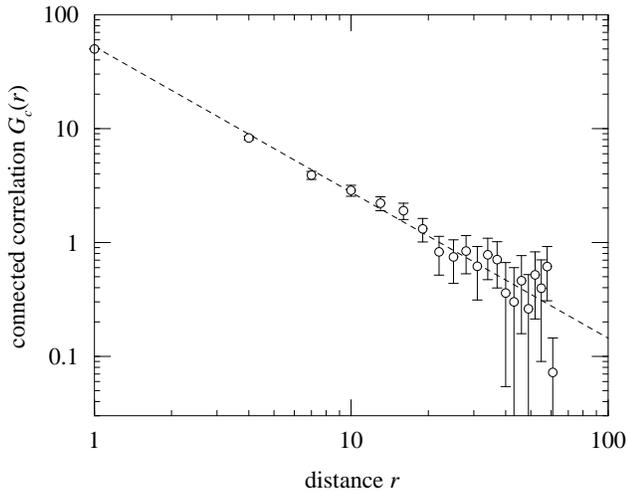,width=3.25in}
\end{center}
\caption{The connected correlation $G_c(r)$ for the staggered magnetization
  on the projective plane.  While the data are not as good as those for the
  torus in Figure~\ref{stagtorus}, we get a compatible value of $\eta =
  1.28 \pm 0.09$ for the correlation exponent.}
\label{stagpp}
\end{figure}

We have also measured spatial correlations in the staggered magnetization
on both the torus and the projective plane (Figures~\ref{stagtorus}
and~\ref{stagpp}, respectively).  The results on the projective plane have
larger statistical errors than those on the torus, because of the need to
stay well away from sites with local curvature in performing the
calculations.  Fits to the data yield values of $\eta = 1.27 \pm 0.01$ on
the torus and $\eta = 1.28 \pm 0.09$ on the projective plane, in reasonable
agreement with theoretical predictions.

\begin{figure}
\begin{center}
\psfig{figure=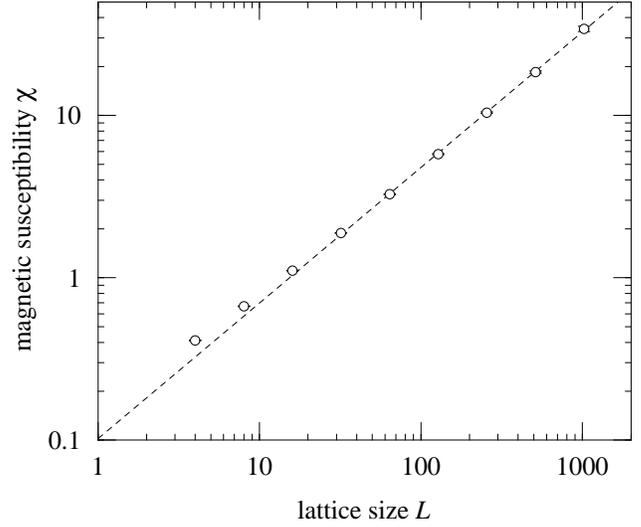,width=3.25in}
\end{center}
\caption{The susceptibility for the (un-staggered) magnetization on the
  projective plane for systems of linear dimension $L$ up to 1024.  The
  dotted line is a fit to the last five points and gives $\gamma/\nu = 0.84
  \pm 0.01$, while theory predicts $\gamma/\nu = 1$.}
\label{susc}
\end{figure}

\begin{figure}
\begin{center}
\psfig{figure=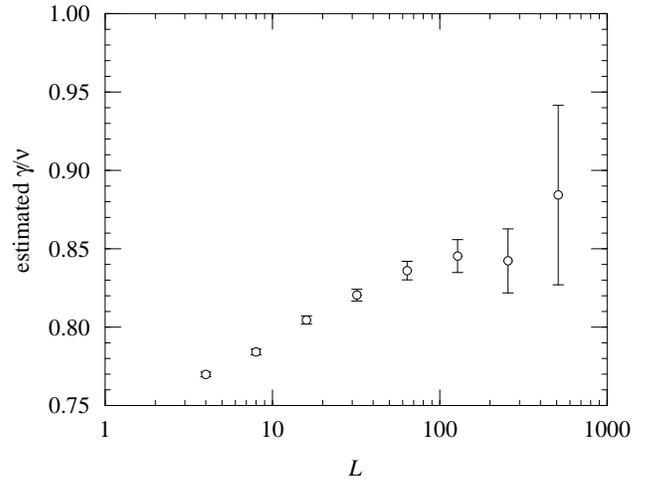,width=3.25in}
\end{center}
\caption{Estimates of $\gamma/\nu$ obtained from fits to simulation
  results for the un-staggered magnetic susceptibility on lattices of size
  ranging from $L$ up to $1024$, as a function of $L$.  These tend upward
  with increasing $L$.}
\label{upward}
\end{figure}

The situation is a little more complicated for our other definition of
magnetization.  Looking at Figure~\ref{map}, we see that the mapping from
heights to spin states has wavelength~2 on the height lattice, which
implies that both the correlation exponent for the spins and the
corresponding susceptibility exponent should be equal to~1.
Figure~\ref{susc} shows our data for the susceptibility on the projective
plane.  A least-squares fit gives $\gamma/\nu = 0.84 \pm 0.01$, which is
some way from the theoretical prediction, but, as Figure~\ref{upward}
shows, the value is increasing steadily with system size, so the
discrepancy is again probably due to logarithmic corrections.

\begin{figure}
\begin{center}
\psfig{figure=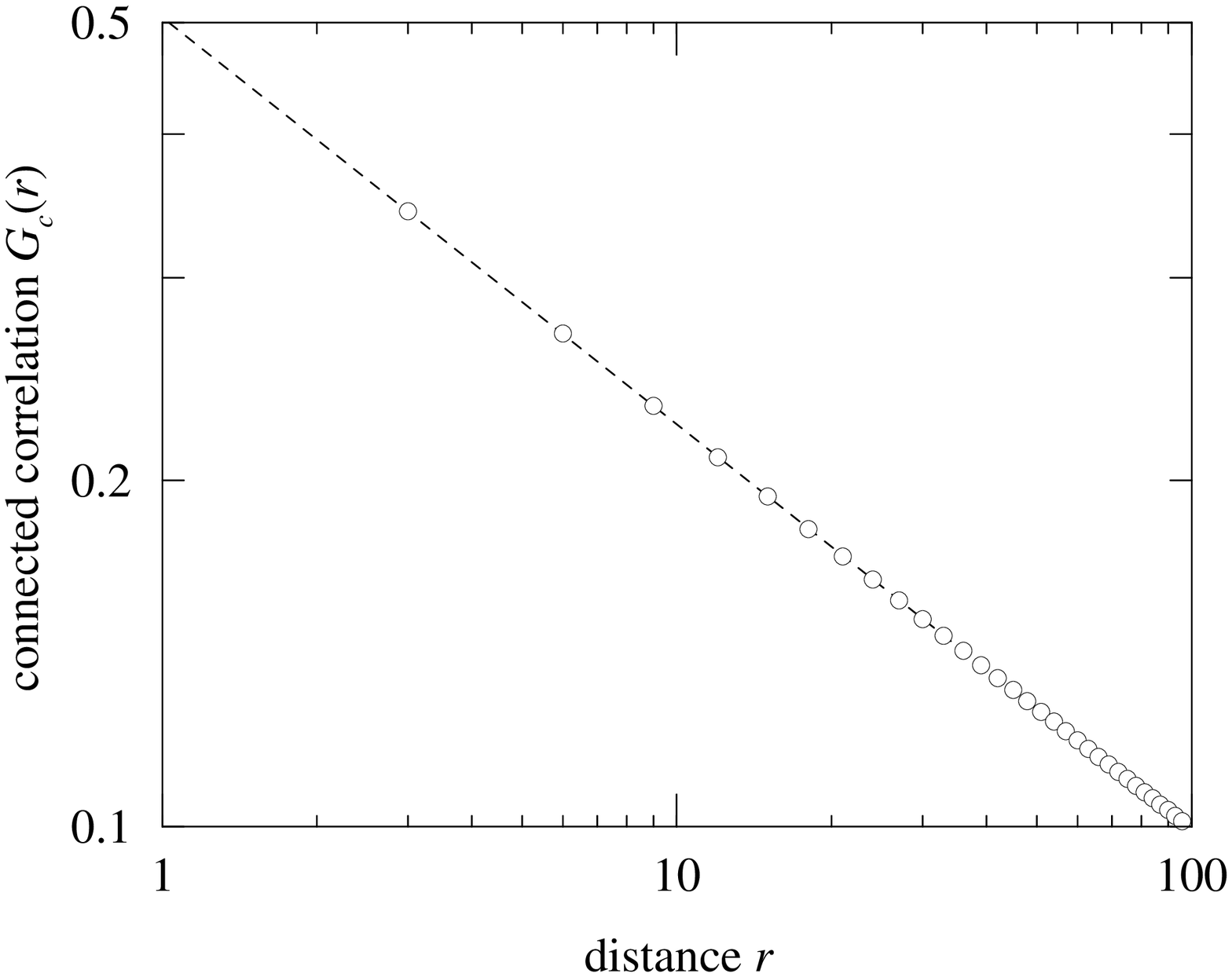,width=3.25in}
\end{center}
\caption{The connected correlation $G_c(r)$ for the (un-staggered)
  magnetization on the torus.  The best power-law fit is indicated by the
  dotted line and gives $\eta = 0.35 \pm 0.01$.  Theory predicts $\eta =
  \frac13$.}
\label{spin}
\end{figure}

However, when we look at the spin--spin correlation function,
Figure~\ref{spin}, we find that $\eta = 0.35 \pm 0.01$, which is nowhere
near~1.  Our values for the exponents thus appear to violate the Fisher
law.  The explanation of this result is as follows.

\begin{figure}
\begin{center}
\psfig{figure=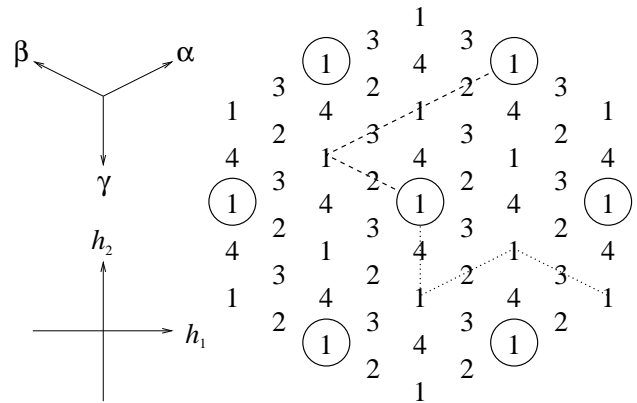,width=3.25in}
\end{center}
\caption{The sublattice of possible height differences between two sites
  along a principal lattice direction, both with spins in state~1, whose
  separation on the lattice is a multiple of~3 times the lattice parameter.
  The dashed path $2\beta + 4\alpha$ is allowed, but the dotted path
  $2\gamma + 2\alpha - 2\beta$ is not, since it contains terms of different
  signs.  The wavelength of the sublattice is $2\sqrt{3}$, which implies
  that $\eta = \frac13$ for the spin--spin correlation function.}
\label{sublattice}
\end{figure}

The correlation function shown in Figure~\ref{spin} is measured, naturally
enough, along one of the three principal directions on the toroidal
lattice.  If two sites lie along such a direction and their spins are in
the same state, then their heights are constrained to particular
sublattices of the height lattice.  For instance, suppose that the distance
between them is a multiple of~3, and that one site has state~1 and height
$(0,0)$.  Then the height of the other site is a sum of multiples of three
times $\alpha$, $\beta$ and $\gamma$, all with the same sign since $\vt$ is
constant along any of the lattice directions.  Since
$\alpha+\beta+\gamma=0$, we can cancel terms in threes until only two kinds
of terms are left, say $\alpha$s and $\beta$s.  Since the number of these
is still a multiple of three, and since $\alpha$ and $\beta$ (or more
precisely, the unit vectors in those directions on the height lattice) have
a component of $\frac{1}{2}$ along the $h_1$ axis, the only heights we can
end up with are ones with $h_1 = 3k/2$ for some integer $k$.  The set of
such heights which correspond to spin state~1 is shown in
Figure~\ref{sublattice}.  More generally, once we choose the distance
between two sites and the state of the spin on one of them, the only
heights the other site can have and still be in the same spin state are
those on one of the three sublattices corresponding to that state.  The
wavelength of each such sublattice is $2\sqrt{3}$, and
Equation~\ref{etaeqn} then gives $\eta = 1/3$, in good agreement with our
simulation results.

This phenomenon, in which some constraint gives an operator a wavelength on
the height lattice longer than that for the corresponding susceptibility,
may apply to other operators and systems like this one.  Therefore, it
appears that the Fisher scaling relation cannot always be applied in a
straightforward fashion.

\section{Conclusions}
\label{secconcs}
In this paper we have studied the four-state Potts antiferromagnet on the
triangular lattice at zero temperature.  By examining the spectrum of
defect types which can appear in the model and identifying the conservation
laws which govern their interactions, we have been able to define a Burgers
vector for the model and thus show that the ground-state ensemble has a
well-defined height representation.  The height is, in this case,
two-dimensional and may be the simplest example to date of a vector height.

Using the height representation, we have been able to demonstrate a number
of results.  We have shown that the model is equivalent to a three-state
Potts antiferromagnet on the bonds of either the triangular or hexagonal
lattice, or on the sites of the Kagom\'e lattice, that pairs of defects
feel entropic forces between them in proportion to the dot product of their
topological charges, and that the spin--spin correlations in the
ground-state ensemble must decay algebraically at large distances.

We have calculated exact values for a variety of critical exponents.  The
scaling exponent $\eta$ for the spin--spin correlation function is of
particular interest because the wavelength of the spin operator on the
height lattice turns out to be longer than the fundamental periodicity of
the height-to-spin mapping due to a constraint on paths connecting sites
along a principal lattice direction.  This gives a value of $\eta =
\frac13$ even though the corresponding susceptibility exponent $\gamma/\nu
= 1$---an apparent violation of the Fisher scaling relation.

We have also used the height representation to prove for the first time
that the Wang-Swendsen-Koteck\'y cluster Monte Carlo algorithm is ergodic
for the four-state model on the triangular lattice.  Our proof however
requires that the lattice have a topology which possesses no
non-contractible loops of infinite order, and this means that simulations
on the torus (which has such loops) are probably not ergodic, calling
previous simulations of this and related models into question.  Simulations
on lattices with free boundary conditions or lattices with the topology of
the sphere or the projective plane are, on the other hand, provably
ergodic, and we have performed extensive simulations on the projective
plane.  We find reasonable agreement between the values of the critical
exponents measured in these simulations and both theoretical values and
values from previous numerical studies.  There are, however, significant
logarithmic corrections to scaling associated with the fact that the height
model is exactly at its roughening transition, and this means that we have
to go to extremely large lattice sizes to see the expected behavior.

\section*{Acknowledgements}
We wish to thank Alan Sokal and Jesus Salas for sharing with us their
observation that the WSK algorithm is not ergodic on lattices with toroidal
boundary conditions, and Michael Lachmann for useful discussions.  C.M.
also thanks Molly Rose and Spootie the Cat for their support.

{\em Note added.} During the course of these investigations, we learned of
an unpublished manuscript by Henley which also shows that the four-state
antiferromagnet studied here is equivalent to the three-state one on the
Kagom\'e lattice.

\end{document}